# How sustainable are different levels of consciousness?

Erik J Wiersma

Affiliation:  Alba Biologics, Toronto, Ontario, Canada,

Email: erikwiersma@rogers.com

Abstract:

*The human brain processes a wide variety of inputs and does so either consciously or subconsciously. According to the Global Workspace theory, conscious processing involves broadcasting of information to several regions of the brain and subconscious processing involves more localized processing. This theoretical paper aims to expand on some of the aspects of the Global Workspace theory: how the properties of incoming information result in it being processed subconsciously or consciously; why processing can be either be sustained or short-lived; how the Global Workspace theory may apply both to real-time sensory input as well as to internally retained information.*

*This paper proposes that: familiar input which does not elicit intense emotions becomes processed subconsciously and such processing can be continuous and sustained; input that elicits relatively intense emotions is subjected to highly sustainable conscious processing; input can also undergo meta-conscious processing. Such processing is not very sustainable but can exert control over other cognitive processes. This paper also discusses possible benefits of regulating cognitive processes this way.*

**Keywords:** *Consciousness, Global Workspace theory*, *cognition, emotion*

# 1.    Introduction

*1.1 The Global Workspace Theory*

Consciousness is a multi-faceted concept, and this paper is limited to some aspects; to access consciousness (the type that can be used for reasoning and directing acting) but not phenomenal consciousness (subjective sensations such as the experience of colours, tastes, smells etc.). Also, this paper will look at different levels of wakeful consciousness, especially how such levels are attained and sustained. A defining feature of consciousness is that a person can report on the content of what is being processed; in contrast, the content of subconscious processing is not reportable. These two levels (consciousness and subconsciousness) can, as presented later in this paper, be divided into other levels.

According to the Global Workspace (GW) theory (Baars, 1997; Baars 2005; Baars, Franklin & Ramsoy, 2013) consciousness involves the broadcasting of selected information to many different brain regions. During the onset of conscious processing there is a shift from isolated local processing towards shared, global processing. Broadcasting is facilitated by the thalamus supporting a cortical 'up state', enabling different cortical regions to communicate with one another in a synchronized fashion (Rigas & Castro-Alamancos, 2007; Steriade, McCormick & Sejnowski, 1993). Cortical broadcasting leads to extensive cognitive resources being spent on one (out of many) stream of information so that it is processed accurately and effectively. Ongoing competition for mental resources determines which stream of information will be broadcasted. When information is processed consciously it is often perceived as being unitary and internally consistent — this is likely a result of utilizing substantial resources on focussed processing. Over time the brain's focus will shift to different streams of input (James, 1890, p. 243; Hasenkamp, Wilson-Mendenhall, Duncan & Barsalou, 2012; Logothetis, Leopold & Sheinberg, 1996; Vanhaudenhuyse et al., 2010). This ability, to re-direct the brain's cognitive resources, is useful when responding to a changing environment.

There is experimental support for GW theory; for example, experiments utilizing functional magnetic resonance imaging indicate that conscious processing coincides with integration among different brain regions (Fahrenfort et al., 2012; Godwin, Barry & Marois, 2015). Also, it has been possible to dissect the onset of conscious processing (Dehaene, 2014, pp. 121-142) both in terms of stimuli thresholds that need to be exceeded, as well as tracking the time course of how different brain regions become successively involved during the onset of consciousness. Studies provide support for three levels of cognitive processing (Dehaene, Changeux, Naccache, Sackur & Sergent, 2006; Dehaene, 2014, pp.190-193). The lowest level, subliminal processing, is caused by stimuli having weak bottom-up strength. Such stimuli do not result in wide broadcasting nor the ability to report on the stimuli. The next level, pre-conscious processing, is caused by strong stimulus in the absence of strong top-down attention. It has the potential of becoming reportable, albeit after a temporal delay. The third level, conscious processing, is caused by strong stimulus strength and strong top-down attention, and it is reportable. Two publications (Zylkerberg, Slezak, Roelfsema, Dehaene & Sigman 2010; Zylkerberg, Dehaene, Roelfsema & Sigman, 2011) describe reverse engineering where these levels of brain processes were modeled by a computer software. This software was able to accurately replicate some aspects of short-term processing of simple visual stimuli. This model, described in three publications (Dehaene, Changeux, Naccache, Sackur & Sergent, 2006;

Zylkerberg, Slezak, Roelfsema, Dehaene & Sigman, 2010; Zylkerberg, Dehaene, Roelfsema & Sigman, 2011), will be referred to as 'the Router model'.

Another GW-based computer model, LIDA (reviewed by Franklin et al., 2016), also has three levels of cognitive processing, and incoming sensory input is routed to any of these levels through their degree of salience (Franklin & Baars, 2010). High level processing, which is accurate but slow, involves a greater number of cycles of repeat processing than do lower levels of processing (Faghihi, Estey, McCall & Franklin, 2015). This feature, to undergo consecutive and iterative rounds of processing, contrasts with the Router model's short-term start-to-finish processing. Repeated, sustained processing is an important step towards replicating human-like cognition. Humans often perform consecutive rounds of cognitive processing to solve problems (Dehane & Sigman, 2012), each round lasting hundreds of milliseconds. Also, some aspects of conscious experiences require integration of input received at different points in time, i.e. carry-over of information from one round of cognitive processing to the next. One such aspect is that a sense of continuity, such as continuous motion, is obtained by integrating sensory information from different time points (Geldard & Sherrick, 1972; Kolers & von Grünau, 1975). Another aspect is that temporal integration enables patterns to be recognized; it has been claimed that the ability to recognize patterns in a long auditory sequence is a marker of conscious processing (Bekinschtein et al., 2009). An additional aspect is that temporal integration is important for the richness of a conscious experience. Our visual experience of the environment is often richer than the information we perceive at any given instance and this because of temporal integration (Melchers & Morrone, 2007). Taken together, integrating information from sustained, repeated cycles of cognitive processing appears to be important for several attributes of human consciousness.

This paper seeks to build on the Router and LIDA models to further our understanding of some aspects of consciousness, especially the sustainability of processing. In order to launch such ideas, the remainder of this Introduction will present a brief overview of some principles of cognitive processing and emotions.

*1.2 Sensory-coupled Cognition*

The brain can rapidly and accurately process a diverse and vast amount of information. This involves both real-time processing of input sensory organs, which is briefly summarized in this section, as well as processing that is not directly coupled to sensory input, discussed in the next section. Efficient processing of sensory input is made possible by several operating principles, including the following:

The brain uses a high degree of parallel and distributed processing during initial steps of cognition (Cisek & Kalaska, 2010; Feldman & Ballard, 1982; Nassi & Callaway, 2009). Such a principle, which operates at different anatomical scales of the brain, allows for efficient processing. At the smaller scale, parallel processing is carried out by a large number of neural processors that are arranged in series and in layers (Hinton, 2007; Poggio, 2016). Such serial layers form local networks with signalling among adjacent layers.

Parallel processing also operates at the brain's larger anatomical scale. Distinct brain regions carry out primary processing for different sensory streams, e.g. there are separate centres for visual, auditory and olfactory inputs. Such brain regions, as well as more domain-general regions, are connected through network hubs (Bola & Sabel, 2015; Moon, Lee, Blain-Moraes, & Mashour, 2015; Van Den Heuvel & Sporns, 2013). Some hubs confer regional, short-range connectivity. Other, 'rich club' hubs, connect a larger number of brain regions over longer physical distances, and these are seen as being important for large-scale integration of information.

Another approach for effective cognition, related to broadcasting of information, is the ability to focus processing on some incoming information at the expense of other, i.e. selective attention (Dehaene, 2014, p.21-22; Yantis, 2008). Attention involves competitive interactions during inter-neuronal communications such that signalling from some neurons but not others are forwarded for processing (Moran & Desimone, 1985). Attentional mechanisms are at work during several stages of cognitive processing, both in the prefrontal cortex (Kim, Ährlund-Richter, Wang, Deisseroth, & Carlén 2016) and in other brain regions (Rueda & Posner, 2013).

Effective processing is also made possible by memories of previous experiences. Creating rich associations is an integral part of forming memories as well as in retrieving them. Memories can be formed more efficiently when items are presented in a context, such as linked to imagery (Groninger, 1971) or in an emotional context (Yesavage, Rose & Bower, 1983) as compared to an absence of such contexts. The hippocampus has a key role in forming associations, e.g. in linking a new item or event with information about its time and place (Mankin et al., 2012; Staresina & Davachi, 2009). The hippocampus has neural connections with several regions of the cortex, and such connections are regarded as important for integrated memories. During memory retrieval a single cue, such as a smell (Gottfried, Smith, Rugg. & Dolan 2004), spatial cue (Karlsson & Frank, 2009) or words (Horner, Bisby, Bush, Lin, & Burgess, 2015) can bring back larger episodic memories. Such cue-based recall of memory helps the processing of incoming sensory information.

*1.3 Sensory-decoupled cognition*

In addition to sensory-dependent cognition, described in the previous section, cognitive processing can also involve content that is unrelated to immediate sensory input. Such sensory-decoupled processing (hereafter called 'decoupled cognition') includes activities such as day dreaming, mind wandering, creative thinking and rumination (Christoff, Irving, Fox, Spreng & Andrews-Hanna, 2016). Studies (Killingsworth & Gilbert, 2012; Klinger & Cox 1987; Song & Wang, 2012) suggest that 20-50% of our time awake is spent mind wandering.

Decoupled cognition is not just abundant, it can also lead to important outcomes: planning for the future and achieving future goals (Stawarczyk, Majerus, Maj, Van Der Linden, & Argembeau, 2011; Smallwood, Ruby & Singer, 2013), reprocessing of memory (Wang et al. 2009) as well as creativity and problem solving (Beaty, Benedek, Silvia, & Schacter, 2016; Ritter & Dijksterhuis, 2014; Sio & Ormerod, 2009). Decoupled cognition can also have negative effects such as poorer performance of ongoing tasks (He et al., 2011; Stawarczyk, Majerus, Maj, Van Der Linden & Argembeau 2011) and unhappiness (Killingsworth & Gilbert, 2012). It has been proposed (Allen et al. 2013; Vatansever, Manktelow, Sahakian, Menon. & Stamatakis, 2015) that decoupled cognition has an overall positive impact when it is flexibly balanced with sensory-dependent cognition.

Some brain regions, those of default mode network (Grecius, Krasnow, Reiss & Menon, 2003; Mason et al., 2007), show preferential activation during decoupled cognition, but several other regions appear not to have such a preference. Many brain regions become activated when we plan for or are memorizing an event (decoupled processing) as well as when we experience something in real-time experience and perform motor action (sensory-dependent processing) (Barsalou, 2008, p. 627; Decety & Grèzes, 2006; Nyberg et al., 2000). Because there are similarities between decoupled and sensory-dependent processing it has been suggested (Shanahan, 2006; Song & Tang, 2008) that GW theory applies to both types of processing. This idea will be expanded upon in this paper.

*1.4 Emotion*

Emotion is a central and complex part of consciousness. Although experts have found it difficult to agree on a definition for 'emotion' there is agreement that characteristics of emotion include "recruits response systems" and "motivates cognition and action" (Izard, 2007, p. 271). This paper will focus on such functional characteristics.

Emotions are evoked by internal or external cues and result in different types of responses. A wide variety of functions are influenced by emotions, such as attention (Bocanegra & Zeelenberg, 2011; Fredrickson & Branigan, 2005; Koster, Crombez, Van Damme, Verschuere & De Houwer, 2004; Öhman, Flykt & Esteves, 2001), reasoning and decision making (De Martino, 2006; Blanchette, 2006; Sohn et al., 2015; Zeelenberg, Nelissen, Breugelmans & Pieters et al. 2008) as well as arousal and memory (Bradley et al., 1992; Cahill & McGaugh, 1995). It has been claimed that emotions are economical and effective means to respond to input in a beneficial way (Bechara & Damasio, 2005; Lowe & Ziemke, 2011; Muramatsu & Hanoch, 2005).

Emotions are caused by neural activity and can arise from both sensory-dependent (Bocanegra & Zeelenberg, 2011; Öhman, Flykt & Esteves, 2001) and decoupled processing (Killingsworth & Gilbert, 2012; Ruby, Smallwood, Engen & Singer, 2013). The neural activity that gives rise to emotions can be more or less complex. On one hand, it is understood that the processing that gives rise to somatosensory sensations (e.g. pain and temperature) is less complex (Lloyd, McGlone & Yosipovitch, 2015; Ross, 2011). On the other hand, many other emotions are caused by more complex processing. According to appraisal theories (Roseman & Smith, 2001; Moors, Ellsworth, Scherer & Frijda, 2013) emotions arise from cognitive evaluation of events and situations. It has been proposed (Scherer, 2009; Scherer & Mueleman, 2013) that such evaluations involve appraisal of several categories (relevance, coping potential, normative significance, as well as their subcategories) and that, depending on how such categories score, specific emotions are evoked; joy, rage, fear, sadness. A similar idea, using a different scheme, was proposed by Roseman (2013). These hypotheses have received some experimental support. Scherer and Mueleman (2013) found that subjects' reports of their emotions correlated with the ratings they provided for different appraisal categories. In another study Roseman and Evdokas (2004) was able to alter the emotions that subjects experienced through controlled manipulation of their appraisal categories.

## 2. A Modified Global Workspace Model for Consciousness

### 2.1 Overview of the Model

Figure 1 presents a model for cognitive processing that is based on GW theory and shares some of its elements with the Router model and the LIDA model.

During the first step, sensory input is received by local networks of parallel and layered processors, and these transform the incoming information. Such processing also involves the information being appraised, and may lead to that emotions are evoked. Emotions lead to several types of responses and functions, e.g. to attention and access to short-term memory where information can be stored for several seconds (McGaugh, 2000). This short-term storage enables the information to be carried forward to subsequent rounds of processing. Attention also allows for access to network hubs as well as to long-term memories that can be retrieved and utilized for additional rounds of cognition. When information gains access to hubs, it becomes condensed. This condensation will result in some information being filtered away, i.e. to information remaining unconscious. This processed and condensed information gains access hubs where it is broadcasted and received by local networks. This closes one round of cognitive processing. Provided that the broadcasting is sufficiently widespread we will become conscious of the information.

Figure 1 is a generic model of cognitive processing. The next sections of this paper will discuss different levels of consciousness and proposes how variations (Figure 2) of this generic model describe different levels of processing.

### 2.2 Sustainability of Cognitive Processing

Conscious processing has a bottleneck — generally we are not conscious of several streams of information at the same time (Pashler, 1994; Logothetis, Leopold & Sheinberg, 1996). There are indications (Dixon, Fox & Christoff, 2015) that this bottleneck applies both to sensory-dependent processing and to decoupled processing.

There are reasons for believing that rich club hubs are responsible for the limited capacity of conscious processing, and that this limitation is due to the burden of broadcasting information. Rich club hubs connect brain regions through long-distance axonal projections (Rubinov, Ypma, Watson & Bullmore 2015). Such a mode of wiring is not as economical as utilizing a more distributed connectivity but it brings the advantages of adaptable and integrated processing (Bullmore & Sporns, 2012). Analysis of gene expression also suggests there is cost; the transcriptional signature of rich club hubs indicates high metabolic activity (Fulcher & Fornito, 2016), exceeding that of several other neuronal structures (Vertes et al., 2016), suggesting that rich club hubs operate at a high capacity. If rich club hubs are indeed the bottleneck of conscious processing, then it becomes important to select the streams of information that gain access to these hubs. This paper suggests that it is emotions that create such access. One of the functions of emotions is selective attention (Bocanegra & Zeelenberg, 2011; Fredrickson & Branigan, 2005; Koster, Crombez, Van Damme, Verschuere & De Houwer, 2004; Öhman, Flykt & Esteves, 2001) which, in context of this model, is understood as creating access to the global workspace through rich club hubs. Other GW-based models make similar proposals, that saliency, feelings (Franklin, Madl, D'Mello & Snaider, 2014, Franklin et al. 2016) and affect (Shanahan, 2005) enable information broadcasting and repeat processing.

In the proposed model (Figure 1) local networks of processors create emotions, allowing rich club hubs to broadcast information, and enabling it to undergo repeated rounds of processing. Different streams of information are processed at any given time by different local networks, and unrelated streams compete for access to the same rich club hubs. It is proposed that a given stream can be sustained only if it generates sufficient emotion to be broadcast to a large enough number of local networks. If the number of networks that are engaged in repeat processing cannot be maintained, then that stream of processing will either stop or become diminished. In other words, it is proposed that sustainability is the amount of emotional intensity generated during processing compared to the cognitive effort required for such processing to continue, and this can be expressed as:

sustainability of processing = emotional intensity / cognitive effort

In order for this equation to be experimentally testable, one needs methods for measuring cognitive effort and emotional intensity. Several methods have been described for measuring cognitive effort, including electroencephalography (Antonenko, Paas, Grabner & Van Gog, 2010), pupillometry (Kahnemann & Beatty 1966) and response time measurements (Dux et al., 2009; Smallwood, McSpadden & Schooler, 2007). Also a number of methods (reviewed by Mauss & Robinson (2009)) have been used for measuring different emotions.

Next, sections 2.2 through 2.4 will look at how this equation applies to different levels of consciousness; briefly, it is proposed that different types of sensory input create different levels of emotional intensity and cognitive effort, and this results in different levels of cognitive processing that have different degrees of sustainability. Thereafter, Sections 2.5 and 2.6 will expand these concepts to decoupled processing.

*2.2 The Subliminal Level of Sensory-coupled Processing*

At this lowest level of consciousness, the brain receives sensory input but performs only cursory processing, i.e. input is processed by early sensory areas of the brain (Dehaene, 2014, pp.121-123; Dehaene, Changeux, Naccache, Sackur & Sergent, 2006), but is not forwarded to other regions of the brain. Such input has been described as having low stimulus strength (Dehaene, Changeux, Naccache, Sackur & Sergent, 2006). Here, such input is viewed as not generating emotion (Table 1, row 1) and, as a consequence, not being sustainable. Inability to access rich club hubs mean that a first round of acquired processing cannot be completed (Figure 2, panel 1). Despite such processing not being sustainable, it need not be completely non-productive: subliminal processing has the capability to influence attention to a subsequent, consciously processed stimulus provided that the timing between the two stimuli is very short (Naccache, Blandin & Dehaene, 2002).

*2.3 The Preconscious Level of Sensory-coupled Processing*

Preconsciousness is the next level of consciousness, and it is more capable than subliminal processing. One such capability is to store partially processed information while awaiting for access to conscious processing to become available. Some experimental set-ups (attentional blink and psychological refractory period; reviewed by Dehaene et al. (2006)) provide strong evidence for this capability, and have lead to the view that that preconscious processing is "conscious-in-waiting" (Dehaene, 2014, p. 191). The LIDA model provides a somewhat different view — it proposes that preconscious processing is ongoing and active, and that such activity underpins conscious processing (Franklin & Baars, 2010).

There are reasons to believe that preconscious processing is not limited to "consciousness-in-waiting" or to providing support for conscious processing and that it can also lead to functional outcomes all by itself. A number of activities can be performed subconsciously such as integration of multisensory information (Salomon et al. 2016), social signalling (Lakin & Chartrand, 2003), reading and performing additions (Sklar, Nevy, Goldstein, Mandel, Maril & Hassin 2012), decision making (Galdi, Arcuri & Gawronski, 2008), and playing chess (Kiesel, Kunde, Berner & Hoffmann, 2009). Since some of these processes are relatively complex, it seems reasonable that they involve multiple rounds of cognitive processing, i.e. that preconscious processing can be sustained.

The studies cited above describe mental operations that had been overlearnt through past events, and this is likely a hallmark for this type of preconscious processing. It has been claimed (Bargh, 1997) that such automated processing is very common; that much of our everyday lives relies on learnt processes that are executed subconsciously. Overlearned processes have been well studied. When a cognitive task is performed regularly, it results in changes to neural structure (myelination) and routing (plasticity) (Chevalier et al., 2015; Mensch et al. 2015; O'Rourke, Gasperini & Young, 2014). Such cellular changes can result in neurons responding more selectively and efficiently to specific stimuli (Baker, Behrmann & Olson, 2002). At the macro-anatomical level, training results in reduced activation of fronto-parietal and other brain regions (Chein & Schneider, 2012; Garner & Dux, 2015; Dux et al., 2009). In other words, whereas the performance of a new task often requires wide-spread brain activation, training results in the same task being performed with far less broadcasting of information. The reduction in fronto-parietal activity could either be due to overlearnt processing not being highly dependent on these regions (Kelly & Garavan, 2005) or, alternatively, that extensive practise has led to fronto-parietal activity becoming more efficient (Garner & Dux, 2015; Dux et al., 2009).

The neural changes that occur during learning allows a task to be executed rapidly (Lee, Seo & Jung, 2012) and also to be performed concurrently with other tasks (Garner & Dux, 2015; Dux et al., 2009). This suggests that the cognitive effort for such processing is relatively low. Another important observation is that training often leads to a reduced emotional response (Carretié, Hinojosa & Mercado, 2003; Fisher et al., 2003; Rankin et al. 2010). Assuming that both cognitive effort and emotional intensity are low (Table 1, row 2), one may calculate the sustainability for preconscious processing; as compared to other levels of consciousness (Table 1, rows 1-4), the preconscious level has an intermediate degree of sustainability.

## 2.4 The Conscious Level of Sensory-coupled Processing

Conscious processing is characterized by being reportable, and as discussed above, by being effortful. Conscious processing is also seen as being more capable and flexible than preconscious processing, notably through the greater involvement of the prefrontal cortex (Daw, Niv & Dayan 2005; Eslinger & Grattan, 1993; Karnath & Wallesch, 1992). This paper suggests that it is the degree of emotional intensity that determines whether sensory information will be processed consciously or not. It is commonly recognized that novel stimuli can give rise to intense emotions, and that novel stimuli tend to be processed consciously. However, conscious processing and intense emotions are not limited to novel stimuli; in some contexts, re-exposure to a stimulus results in a sensitised response (Grillon & Davis, 1997; Groves & Thompson, 1970; Richardson & Elsayed, 1998). It is suggested that such familiar stimuli can be processed consciously.

Assuming a conscious processing involves a relatively high emotional intensity and an intermediate level cognitive effort (Table 1, row 3), the sustainability of such processing would be greater than that of any other level of processing. This means that a stream of conscious processing is likely to be highly recursive and able to effectively compete with other streams.

## 2.5 Decoupled Processing: Subliminal, Preconscious and Conscious Levels

In addition to sensory-coupled processing, cognitive processing can also be decoupled from immediate sensory input (Introduction). Such decoupled processing can compete with sensory-dependent processing for mental resources (reviewed by: Kam & Handy, 2013; Smallwood & Schooler, 2006), and therefore, it is suggested that decoupled processing needs to be included as part of the GW theory.

This paper proposes that decoupled processing, like sensory-coupled processing, can have different levels of processing, a point also made by others: Moutard et al. (2015) proposed that decoupled cognition can transition from being subliminal to becoming conscious. Also, it has been proposed (Dixon, Fox & Christoff, 2014) that decoupled processing, similar to sensory-dependent cognition, has a lower level of processing that is not resource-demanding, as well as a higher level of processing that requires more resources. There is evidence that memories can induce decoupled processing (Ellamil et al. 2016). This paper suggests that processing of memories, similar to processing of real-time sensory information, can have different emotional intensity, cognitive effort, and sustainability, and that such different parameters determine the cognitive level at which memories are be processed.  More specifically, it is proposed that the relationships outlined in Table 1 apply not only to sensory-dependent cognition but also to decoupled cognition.

The subliminal level of processing is not reportable; also it is short-lived and contained within local networks of the brain (Dehaene, Changeux, Naccache, Sackur & Sergent, 2006).

There is reason to believe that some decoupled activity is subliminal, i.e. in the absence of known sensory input, there is temporary activation in isolated areas and small-world networks of the brain (He et al., 2009; Smith et al. 2012). This paper interprets decoupled subliminal activity as being the activation of memories that do not have a significant emotional content, and therefore unable to complete a first round of cognitive processing (Figure 2, panel 5)

Preconscious processing is also not reportable, and involves a modest degree of integration of different brain regions (Dehaene, Changeux, Naccache, Sackur & Sergent, 2006). There are indications that decoupled processing can be preconscious. After one has formulated a problem there may be an 'incubation' phase before a solution is found. During this 'incubation phase' (Hamard, 1954, pp.13-15; Sio & Ormerod, 2009; Ritter & Dijksterhuis, 2014), we are not conscious of trying to solve a problem. However, the fact that we can suddenly reach insight out of the blue has been taken as an indication of subconscious processing. The literature has examples of complex tasks being processed subconsciously, e.g. in higher mathematics, and this would presumably require multiple rounds of processing. Figure 2, panel 6 proposes how this processing may occur.

Conscious processing is reportable and involves a high degree of integration among different brain regions (Dehaene, Changeux, Naccache, Sackur & Sergent, 2006). Decoupled consciousness, which is not constrained by real-time sensory information, has the potential of being highly flexible and dynamic. It is understood that an ongoing chain of decoupled thoughts is linked together through memories that share some aspects of their content (James, 1890; Gabora & Carbert, 2015). Such serial linkages are important for creative thinking and problem resolution (Beaty, Benedek, Silvia & Schacter, 2016; Ritter & Dijksterhuis, 2014; Sio & Ormerod, 2009).

How can conscious decoupled processing be understood in terms of the model proposed earlier in this paper (Figure 2 and Table 1)? Although this type of processing has different input than sensory-dependent processing (i.e. memories versus sensory input), there are indications that it, like sensory-dependent processing, involves high emotional intensity (Carr & Nielsen, 2015; Foulkes & Fleisher, 1975; Tusche, Smallwood, Bernhardt & Singer, 2014). Also, it can be highly sustainability since it can interfere with conscious sensory-dependent processing (Kam & Handy, 2013; Smallwood & Schooler, 2006). Based on this reasoning, it is proposed that both decoupled consciousness and sensory-dependent consciousness have high emotional intensity and high sustainability (Table 1, row 3). Figure 2, panels 3 and 7 illustrates these two types processing as being very similar albeit that decoupled processing does not utilize real-time sensory input.

*2.6 Decoupled Processing: the Meta-conscious Level*

Meta-consciousness is seen as the highest level of processing, and it is reportable. Whereas other levels of consciousness involve processing (or reflecting) on items, events or concepts, meta-consciousness involves reflecting on thinking processes themselves. Schooler (2002, p. 339) defines meta-conscious[1] processing as "intermittent explicit re-representations of the contents of consciousness". There are several implications of this definition:

First, in order to re-represent consciousness, meta-conscious processing must occur after conscious processing – a 'temporal dissociation' (Schooler 2002). This means that meta-consciousness is decoupled from sensory input (Figure 2, panel 8) and that there is no sensory-coupled level of meta-consciousness (Figure 2, panel 4).

Second, if meta-consciousness processing is the re-representation of consciousness then it is likely to pose a greater cognitive effort than does conscious processing. It has been found (reviewed by Schooler, 2002) that monitoring conscious content results in loss of information, consistent with the idea that meta-conscious monitoring demands more mental resources than does conscious experience. Another indication that meta-consciousness involves a high cognitive effort comes from studies of how it interferes with performance of an unrelated task. It was found (Smallwood, McSpadden & Schooler, 2007) that mind wandering with awareness leads to longer response time of an unrelated task than does mind wandering without awareness.

Third, it may be expected that re-representation (meta-consciousness) may lead to emotions becoming less vivid as compared to when they are directly experienced (consciousness). Experiments indicate that this can indeed be the case (Papies, Pronk, Keesman & Barsalou 2015; Schooler, Areily & Loweenstein, 2003, pp. 56-59; Shapira, Gundar-Goshen & Dar, 2013).

If meta-consciousness involves less emotional intensity and a greater cognitive effort than does consciousness, then it will, according to formula presented earlier, be less sustainable than consciousness (Table 1, rows 3 versus 4). There are indications that meta-conscious processing is indeed less sustainable than other forms of decoupled conscious processing: alcohol consumption (Sayette, Reichle & Schooler, 2009), craving for cigarettes (Sayette, Schooler &, Reichle, 2010) and sleep deprivation (Poh, Chong & Chee, 2016) reduce the proportion of mind wandering that involves awareness. Since these different experimental conditions had a greater impact on the occurrence of meta-consciousness than on decoupled consciousness, it is deemed that meta-consciousness is less sustainable than decoupled consciousness.

There is evidence that meta-consciousness can be productive, i.e. that the act of observing our conscious thoughts can influence how we think, feel and the decisions we make. Hasenkamp et al. (2012) found that the act of catching oneself mind wandering is followed by attention being brought back to task performance, and task performance being resumed. Such an observation is consistent with meta-consciousness exerting regulation on other cognitive processes. In another example, Papies et al. (2015) found that observing one's reactions to pictures of attractive food leads to changes in how subjects rated the attractiveness of food. Also, in a separate extension of the same experiment, it was found that the subjects' choice of food purchase changed.

Others have also argued that high-level processing can have productive outcomes. It has been proposed that executive control is exerted by high-level processes that re-represent / reflect on input (Son & Schwartz, 2002; Zelaso, 2015) and that are decoupled from sensory input (Stanovich, 2009).

---

[1] Some publications use the term 'mind wandering with awareness' rather than 'meta-consciousness', and this paper will use these terms interchangeably.

In summary, it appears that meta-consciousness processing has dual properties. On one hand, it is not very sustainable and, on the other hand, once meta-conscious processing does occur, it has the ability to influence many other mental processes.

## 3. Discussion

This theoretical paper builds on the GW theory proposed by Baars (Baars, 1997; Baars, 2005; Baars, Franklin & Ramsoy, 2013) and elaborated by Dehaene (Dehaene, Changeux, Naccache, Sackur & Sergent, 2006; Zylkerberg, Slezak, Roelfsema, Dehaene. & Sigman 2010; Zylkerberg, Dehaene, Roelfsema & Sigman, 2011), Franklin (Franklin, Madl, D'Mello & Snaider, 2014; Franklin et al. 2016) and their respective colleagues. The model presented in this paper elaborates on how well different levels of consciousness are sustained. Sustained processing is important for solving complex problems and for creating a rich, dynamic understanding of our surroundings (Introduction). In addition, this paper elaborates on how decoupled processing can be incorporated into the same general model as sensory-dependent processing. Such a common framework might simplify and help provide a coherent understanding of these two types of processing.

The proposed model has its limitations. Currently it is a relatively simple conceptual model and, unlike the Router and LIDA models, has not yet been translated into an elaborate computer architecture. Also, additional experiments are needed to corroborate or refute the model, particularly for subliminal and preconscious decoupled processing. The model proposes that each level of processing has a distinct value for emotional intensity, cognitive effort and sustainability (Table 1). It is possible that once more experimental data is available that some heterogeneity is discovered within each level of processing.

If the presented model is essentially correct, then what would be the utility for regulating the different levels of processing as described? There is a trend when comparing the three lowest levels of consciousness, subliminal, preconscious and conscious processing: for each higher level of processing the input receives increasingly higher emotional appraisal, leading to greater access to the global workspace and to higher sustainability of processing. This trend appears to make sense – that large amounts of mental resources are dedicated to tasks that (through emotional appraisal) are deemed to be highly relevant and important, and that the processing of such tasks becomes sustainable. This paper proposes that such a principle applies to both sensory-dependent and to decoupled processing; that allocation of resources applies not just to how we process large amounts of information about a real-time situation (sensory-dependent processing), but also to how we plan for future actions and how we solve problems off-line by retrieving information from the brain's vast memory banks (decoupled processing)

Peculiarly, the highest level of processing, meta-consciousness, does not fully conform to the trends seen for the three lower levels of processing (Table 1). Emotional intensity, cognitive effort and sustainability increase in the order subliminal < preconscious < conscious processing. However, only cognitive effort increases when comparing the highest two levels, conscious <

meta-conscious. The other two parameters, emotional intensity and sustainability, are lower at the highest level of processing, meta-conscious < conscious.

What would be the utility of meta-consciousness having less sustainability than consciousness? This may relate to the idea that meta-consciousness has the ability to regulate other cognitive processes. If the primary utility of meta-consciousness was to exert a regulatory function, and it had no direct role in processing input, it may be best if it was not to too sustainable; if meta-consciousness was brief, then most of our mental resources could be spent on actual processing of input, and only limited resources be spent on regulating such processing.

**Table 1: Relationships between properties of input and cognitive processing**

| Row number | Level of consciousness | Properties of input and how it is processed | | | Sustainability of processing |
| --- | --- | --- | --- | --- | --- |
| | | Familiarity | Emotional intensity | Cognitive effort | |
| 1. | Subliminal | Low, medium or high | Negligible (0) | N/A | N/A |
| 2. | Preconscious | High | Low (1) | Low (1) | 1 |
| 3. | Conscious | Low, medium or high | High (4) | Medium (2) | 2 |
| 4. | Meta-conscious | Medium or high | Medium to high (3) | High (4) | 0.8 |

This table illustrates how different properties of input, and how such input is processed, leads to different degrees of sustainability of processing. This table applies to both sensory input (rows 1-3) as well as to decoupled input (memory) (rows 1-4). The three middle columns illustrate that input can have different degrees of familiarity, evoke different degrees of emotions and impart different cognitive efforts. Each of these different variables is ranked from negligible to high, corresponding to ordinal values ranging from zero to four. The sustainability of processing (right-most column) is calculated; *sustainability of processing = emotional intensity / cognitive effort*.

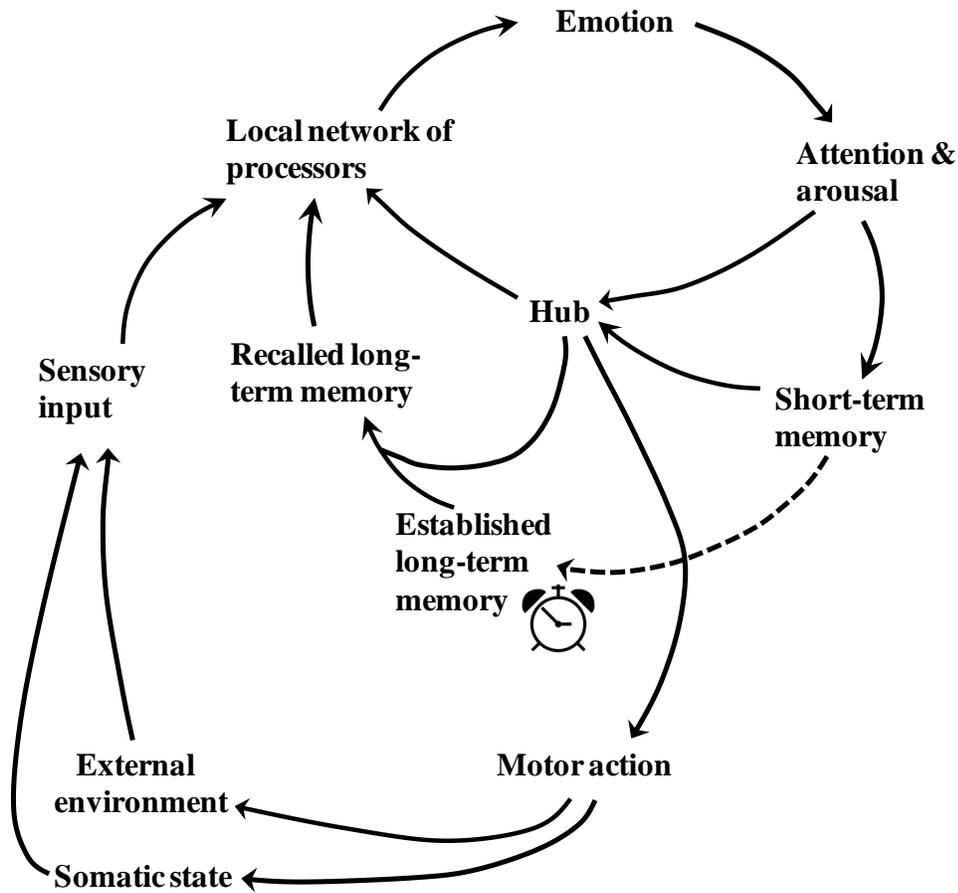

Figure 1:  A proposed workflow model for cognitive processing

The figure illustrates how the different components of cognition interact to form feed-back loops.  The broken arrow and the clock indicate that it takes much longer to create 'established long-term memory' than to complete other parts of the cognitive cycle.

Next page

Figure 2: Proposed workflows during different levels of consciousness

The panels in this figure illustrate different variants of the cognitive cycle that are proposed to occur during different levels of consciousness.  Each panel is a simplified version of the workflow presented in Figure 1, excluding 'motor action', 'external environment' and 'somatic state'.  The processes illustrated in panels 1 and 5 do not form a cognitive cycle.  The processes in panels 2 and 6 form cognitive cycles; however, they do not include formation of new long-term memory, only recall from existing long-term memory. Panels 3, 7 and 8 form cognitive cycles, including forming new long-term memory. These three panels have increased engagement of some parts of their cycles, indicated by larger font (medium activity), and bold large font (high activity); also, these panels have increased broadcasting of information, indicated by two or four arrows originating from 'hub'.

**Sensory-coupled processing**     **Sensory-decoupled processing**

**Subliminal**

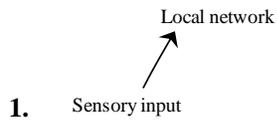

1.

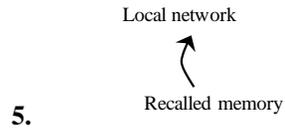

5.

**Preconscious**

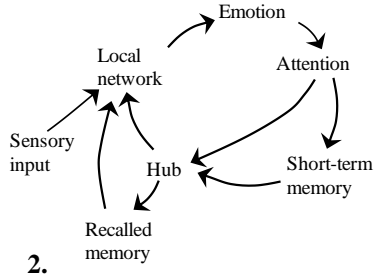

2.

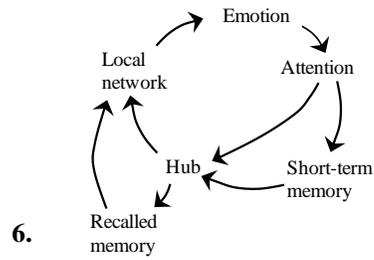

6.

**Conscious**

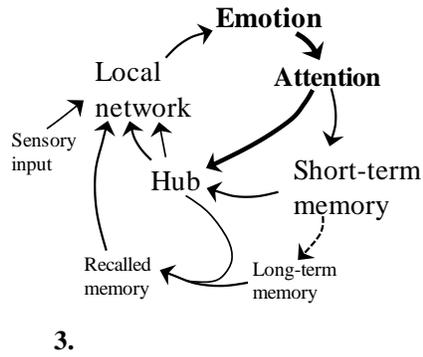

3.

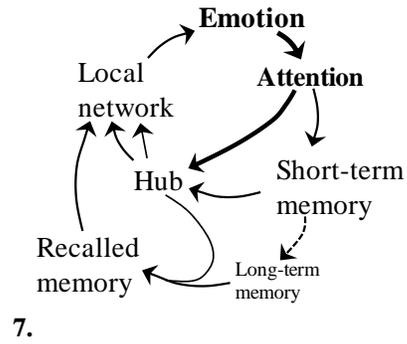

7.

**Meta-conscious**

N/A

4.

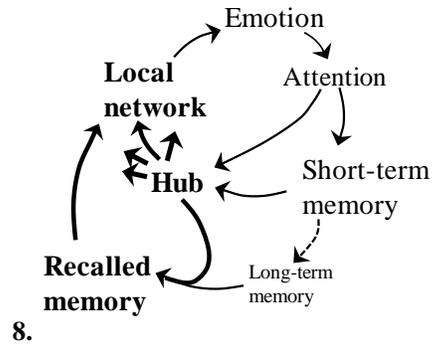

8.

**Acknowledgements:** The author would like thank Drs. Vince Taguchi and Donald Stewart for their valuable comments on this paper

**Funding Details:** This work was not supported by funding

**Disclosure statement:** The author declares no financial conflict of interest in this research